\documentclass{elsart}

\usepackage{amsmath,amssymb,amsfonts}
\usepackage{natbib}
\usepackage{graphicx}% Include figure files
\usepackage{psfrag}

\begin{document}
\bibliographystyle{elsart-num}
\bibliographystyle{unsrt}
\begin{frontmatter}

% Title, authors and addresses

% use the thanksref command within \title, \author or \address for footnotes;
% use the corauthref command within \author for corresponding author footnotes;
% use the ead command for the email address,
% and the form \ead[url] for the home page:
% \title{Title\thanksref{label1}}
% \thanks[label1]{}
\title{An optimum Hamiltonian for non-Hermitian quantum evolution and the complex Bloch sphere}

 \author{Alexander I  Nesterov\corauthref{cor1}}
 \ead{nesterov@cencar.udg.mx}

 \corauth[cor1]{Corresponding author.}
 \address{ Departamento de F{\'\i}sica, CUCEI, Universidad de
Guadalajara, Av. Revoluci\'on 1500, Guadalajara, CP 44420, Jalisco,
M\'exico}
 %\thanks[label3]{}

%\author{}

%\address{}
\begin{abstract}
For a quantum system governed by a non-Hermitian Hamiltonian, we studied the problem of obtaining an optimum Hamiltonian that generates nonunitary transformations of a given initial state into a certain final state in the smallest time $\tau$. The analysis is based on the relationship between the states of the two-dimensional subspace of the Hilbert space spanned by the initial and final states and the points of the two-dimensional complex Bloch sphere.

\end{abstract}

\end{frontmatter}

In view of recent results on optimal quantum evolution and its possible relation with quantum computation and quantum information processing, there is increasing interest in the quantum brachistochrone problem(see e.g., \cite{CHKO,BDHD,BDC,BBJ} and references therein). The  problem consists of finding the shortest time $\tau$ to evolve a given initial state $|\psi_i\rangle$ into a certain final state $|\psi_f\rangle$ under a given set of constraints. Using the variational principle, Carlini {\em et al} \cite{CHKO} has shown that a Hermitian Hamiltonian $\tau$ has a nonzero lower bound. Later, Bender {et al} \cite{BBJ} demonstrated that, for non-Hermitian $\cal PT$-symmetric quantum systems, the answer is quite different, and the evolution time $\tau$ can be made arbitrary small, despite the fact that the eigenvalue constraint is held fixed as it is for the corresponding Hermitian system.

Recently, Brody and Hook \cite{BDHD} have considered the same problem as in \cite{CHKO} using the symmetry properties of the quantum state space and without employing a variational calculus. The purpose of this paper is to extend the approach of Brody and Hook to quantum systems governed by non-Hermitian Hamiltonians (for details and discussion of non-Hermitian physics see, e.g., \cite{WJ,BH,BHC,BS,TT,B}).  We explore the non-Hermitian time-optimal evolution problem from the geometric viewpoint, using the relationship between the states of two-dimensional subspace of the Hilbert space $\cal H$, spanned by $|\psi_i\rangle$ and $|\psi_f \rangle$, and the points of two-dimensional complex sphere (the so-called complex Bloch sphere).

In the quantum brachistochrone problem, finding the shortest possible time requires only the solution of a two-dimensional problem.  Namely, it requires finding the optimal evolution time for the quantum system governed by the effective Hamiltonian acting in the subspace spanned by $|\psi_i\rangle$ and $|\psi_f\rangle$ \cite{BDHD,BDC,BBJ}.
Let the set $\{|\psi_i\rangle,|\psi_0\rangle, \langle \widetilde\psi_i|, \langle \widetilde\psi_0|\}$ form the bi-orthonormal basis of the two-dimensional subspace spanned by the states $|\psi_i\rangle$ and $|\psi_f\rangle$:
\begin{align}\label{Eq1}
 \langle \widetilde\psi_i|\psi_i\rangle=\langle \widetilde\psi_0|\psi_0\rangle= 1, \quad {\rm and}\quad \langle \widetilde\psi_i|\psi_0\rangle =\langle \widetilde\psi_0|\psi_i\rangle= 0.
\end{align}
Then the final state can be written as
\begin{align}\label{B2}
|\psi_f\rangle = a|\psi_i\rangle  + b|\psi_0\rangle,\;
\langle\widetilde\psi_f| = \tilde a\langle\widetilde\psi_i|  + \tilde b\langle\widetilde\psi_0|,
\end{align}
and a generic non-Hermitian Hamiltonian $ H$ acting in the two-dimensional invariant subspace of the Hilbert space $\cal H$ reads
\begin{align}
\label{eqH1}
H=& {\lambda_{0}}(|\psi_i\rangle\langle\widetilde\psi_i| + |\psi_0\rangle\langle\widetilde\psi_0|)+ \frac{\Omega}{2}\cos\theta| \Big(
|\psi_i\rangle\langle\widetilde\psi_i| - |\psi_0\rangle\langle\widetilde\psi_0|\Big) \nonumber \\
&+ \frac{\Omega}{2}\sin\theta \Big( e^{-i\varphi}|\psi_i\rangle\langle\widetilde\psi_0| + e^{i\varphi}|\psi_0\rangle\langle\widetilde\psi_i|  \Big ),
\end{align}
$\lambda_0$ being a complex parameter and $\Omega=\lambda_{+} - \lambda_{-}$, where $\lambda_{+}, \; \lambda_{-}$ are complex eigenvalues of the Hamiltonian $ H$. Thus, the time-optimal evolution problem can be formulated as follows: determine the complex angles $\theta,\varphi$ for which, under a given set of eigenvalue constraints, the initial state $|\psi_i\rangle$ evolves into the final state $|\psi_f\rangle$ in the smallest time $\tau$.

Let $|\psi(t)\rangle = u_1(t)|\psi_i\rangle + u_2(t) |\psi_0\rangle$ and $\langle\widetilde \psi(t)| = \tilde u_1(t)\langle \widetilde \psi_i |+ \tilde u_2(t)\widetilde\langle \psi_0|$ be a solution of the Schr\"odinger equation and its adjoint equation
\begin{align}
i\frac{\partial }{\partial t}|\psi(t)\rangle = H|\psi(t)\rangle, \quad
-i\frac{\partial }{\partial t}\langle\widetilde\psi(t)| =
\langle\widetilde\psi(t)|H\label{S2}
\end{align}
where we have chosen the units with $\hbar =1$. As can be easily shown, the vectors
\begin{equation}
|u(t)\rangle= u_1(t)|u_{\uparrow}\rangle + u_2(t)|u_{\downarrow}\rangle\quad \langle \tilde u(t)|= \tilde u_1(t)\langle u_{\uparrow}| + \tilde u_2(t) \langle u_{\downarrow}|,
\label{S1a}
\end{equation}
where $|u_{\uparrow}\rangle= \left(
  \scriptsize\begin{array}{c}
  1 \\
 0 \\
  \end{array}\right)$
and
$|u_{\downarrow}\rangle= \left(
  \scriptsize\begin{array}{c}
  0 \\
 1 \\
  \end{array}\right)$ denote the up/down states, respectively, satisfy
the following Schr\"odinger equation:
\begin{align}\label{S_3}
i\frac{\partial}{\partial t}|u\rangle = H_{ef}|u\rangle, \quad  -i\frac{\partial}{\partial t}\langle \tilde u(t)| = \langle \tilde u(t)| H_{ef}.
\end{align}
The effective Hamiltonian entering in (\ref{S_3}) is given by
\begin{equation}
\label{eqH2}
H_{ef}= \left(
  \begin{array}{cc}
    \lambda_0 & 0 \\
   0 & \lambda_0 \\
  \end{array}\right)+ \frac{ \Omega}{2}\left(
  \begin{array}{cc}
    \cos\theta & e^{-i\varphi}\sin\theta \\
   e^{i\varphi}\sin\theta &  - \cos\theta\\
  \end{array}
\right),
\end{equation}
and can be expressed in terms of the Pauli matrices as
\begin{equation}\label{eqH3}
H_{ef}= {\lambda_0} {1\hspace{-.15cm}1}+ \frac{1}{2}\,{\mathbf  \Omega}\cdot \boldsymbol \sigma,
\end{equation}
where  ${1\hspace{-.15cm}1}$ denotes the identity operator, and $\mathbf \Omega=(\Omega\sin\theta \cos\varphi,\Omega \sin\theta \sin\varphi, \Omega\cos\theta)$. Using complex Cartesian coordinates,
\begin{equation}
   X=\Omega\sin\theta \cos\varphi,\;Y=\Omega \sin\theta \sin\varphi, \; Z=\Omega\cos\theta, \; \;X,Y,Z \in \mathbb C,
\end{equation}
where $\Omega= \sqrt{X^2+Y^2+Z^2}$, we obtain
\begin{equation}\label{eqH4}
H_{ef}= {\lambda_0} {1\hspace{-.15cm}1}+ \frac{1}{2}
\left(
  \begin{array}{cc}
    Z& X - iY \\
   X + iY &  - Z\\
  \end{array}
\right).
\end{equation}

It is then easy to show that the complex Bloch vector defined as $\mathbf n = \langle\tilde u|\boldsymbol\sigma|u\rangle$, where $\boldsymbol\sigma$ are the Pauli matrices, satisfies the complex Bloch equation
\begin{align}\label{B1}
    \frac {d \mathbf n}{dt}= \mathbf \Omega \times \mathbf n,
\end{align}
which is equivalent to the Schr\"odinger equation (\ref{S_3}) (see, e.g., \cite{CWZL,CT}). The vector $\mathbf n(t) =(n_1(t),n_2(t),n_3(t))$, being a complex unit vector, traces out a trajectory on the complex 2-dimensional sphere $S^2_c$. In the explicit form, its components in terms of the states (\ref{S1a}) read
\begin{align}\label{B2a}
    n_1= u_1\tilde u_2 + u_2\tilde u_1, \;
    n_2= i( u_1\tilde u_2 - u_2\tilde u_1), \;
    n_3= u_1\tilde u_1 - u_2\tilde u_2.
\end{align}
Hence, we obtain the Bloch vector corresponding to the initial state $|\psi_i\rangle$ as $\mathbf n_i=(0,0,1)$.

Let $\mathbf n_f== \langle\tilde u_f|\boldsymbol\sigma|u_f\rangle $ be the complex Bloch vector associated with the final state $|\psi_f\rangle$, and let $\tau$ be the amount of time required to evolve the initial state $|\psi_i\rangle$ into the final state $|\psi_f\rangle$. Then, using the time dependent solution of Eq.(\ref{B1}),
\begin{align}
&\mathbf n(t) = \left(\begin{array}{l}
\sin\theta \cos\theta(1-\cos\Omega t)\cos\varphi + \sin\theta \sin\Omega t \sin\varphi\\
\sin\theta \cos\theta(1-\cos\Omega t)\sin\varphi-\sin\theta \sin\Omega t\cos\varphi \\
\cos^2\theta + \sin^2 \theta \cos\Omega t
\end{array}\right),
 \label{eq1}
\end{align}
with $\mathbf n_i= \mathbf n(0)= (0,0,1)$ and $\mathbf n_f= \mathbf n(\tau) =(\sin\chi \cos\gamma, \sin\chi \sin\gamma, \cos\chi)$, we find
\begin{align}
\label{PT1b}
&e^{-i\varphi} =e^{-i\gamma}\cot\theta\bigg({ \tan\frac{\chi}{2} - i\sqrt{\tan^2\theta -\tan^2\frac{\chi}{2} }}\bigg),\\
\label{PT1c}
&e^{i\varphi} =e^{i\gamma}\cot\theta\bigg({ \tan\frac{\chi}{2} + i\sqrt{\tan^2\theta -\tan^2\frac{\chi}{2} }}\bigg),\\
\label{PT1a}
&\cos \frac{\Omega\tau}{2}= \frac{\sqrt{\cos^2\frac{\chi}{2} - \cos^2\theta}}{\sin\theta}, \quad \sin \frac{\Omega\tau}{2}= \frac{\sin\frac{\chi}{2} }{\sin\theta}.
\end{align}
This yields
\begin{align}
&\cos(\varphi-\gamma) = \cot\theta \, \tan\frac{\chi}{2} \label{PT5a} \\
&\tan \frac{\Omega\tau}{2}=
\frac{\sin\frac{\chi}{2}}{\sqrt{\cos^2\frac{\chi}{2} - \cos^2\theta}}
\label{PT5}
\end{align}
and we obtain the evolution time as $\tau = |\Psi|$, where
\begin{align}
\label{PT6}
\Psi = \frac{2}{\Omega }\arctan\Bigg(\frac{\sin\frac{\chi}{2}}{\sqrt{\cos^2\frac{\chi}{2} - \cos^2\theta}} \Bigg)
\end{align}
In addition, since $\tau$ is a real positive function, the following
constraint must be imposed: $\arg \Psi =0$. This implies
\begin{align}
\label{PT7}
\arg\Omega = \arg \arctan\Bigg(\frac{\sin\frac{\chi}{2}}{\sqrt{\cos^2\frac{\chi}{2} - \cos^2\theta}} \Bigg)
\end{align}
We note that Eq. (\ref{PT7}) can be considered in different ways. For instance, for a given $\Omega$, it restricts all possible final states to those that have the angle $\chi$ satisfying this equation. In contrast, if the eigenvalue constraint is given by $|\Omega| = \rm const$, then (\ref{PT7}) can be considered as an equation to determine the argument of $\Omega$. Finally, for a given $\Omega$ and $\chi$, the equation (\ref{PT7}) yields an implicit function dependent on the variables $\Re \theta$ and $\Im \theta$, and it should be considered in finding the evolution time.

For a given $\Omega$, using (\ref{PT5}) and (\ref{PT7}), we find that the critical points of  the function $\tau(\theta)$ are defined by the following equations:
\begin{eqnarray}
\label{T1}
&&\frac{2\sin\frac{\chi}{2}\cot\theta}{\Omega \sqrt{\cos^2 \frac{\chi}{2}-\cos^2\theta }} =0 \\
&&\arg\Bigg(\frac{1}{\Omega} \arctan\bigg(\frac{\sin\frac{\chi}{2}}{\sqrt{\cos^2\frac{\chi}{2} - \cos^2\theta}} \bigg)\Bigg) =0
\label{T2}
\end{eqnarray}
Then, from  Eq. (\ref{T1}), we obtain two solutions: $\Re \theta =\pi/2$  ($\Im \theta=0$) and $\Im\theta \rightarrow \pm \infty$. Inserting $\theta=\pi/2$ into Eq. (\ref{T2}), we find $\arg\Omega = \arg\chi$, and, for the second solution, we have $\Re \theta = \pi/2 \pm (\arg \Omega - \arg \sin(\chi/2))$.

\begin{figure}[tbh]
%\begin{minipage}[]{8.5cm}
\begin{center}
%\psfrag{1.25}{\huge $ \tau_p$}
%\psfrag{z}{}
\scalebox{0.3}{\includegraphics{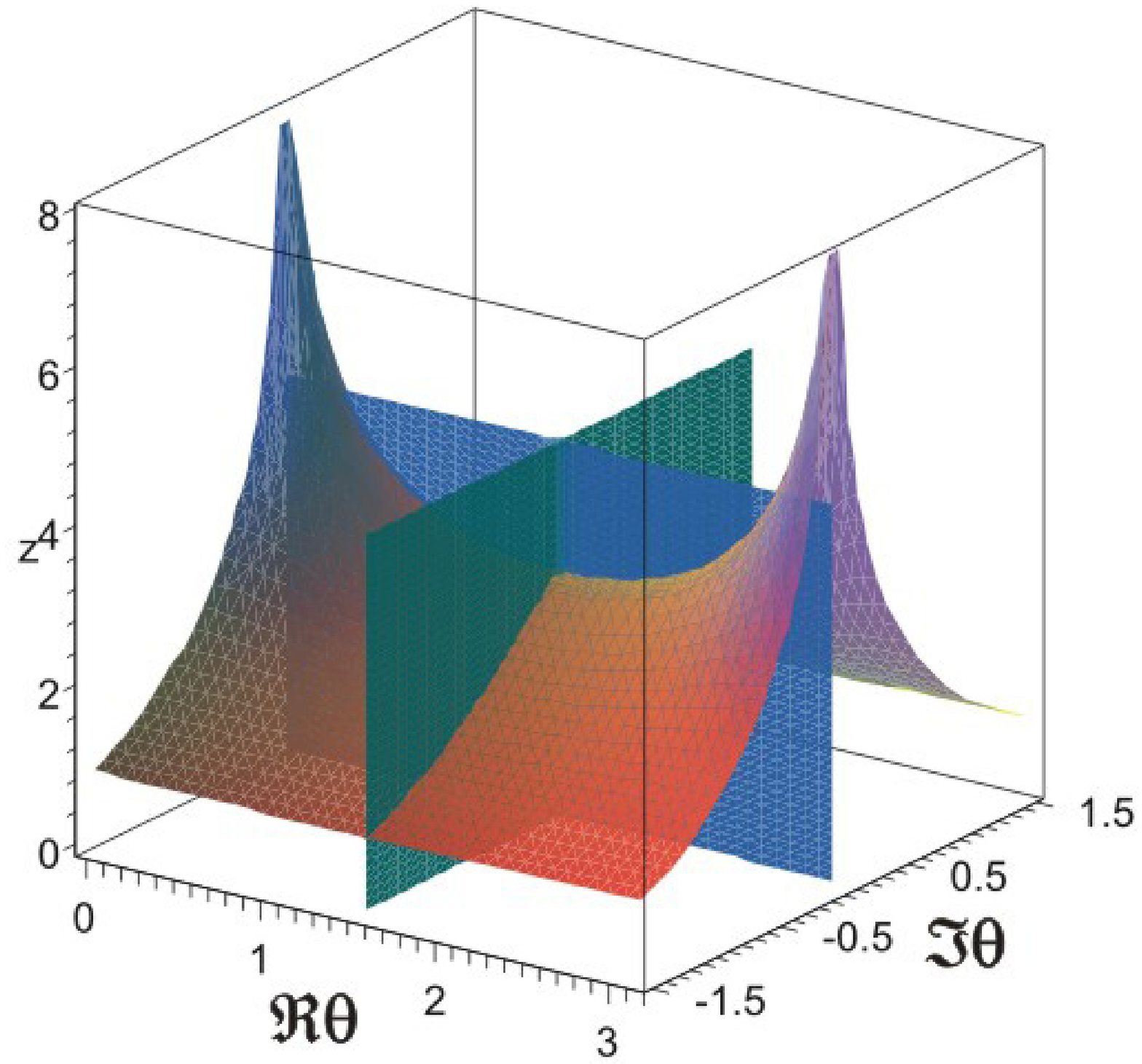}}
\scalebox{0.3}{\includegraphics{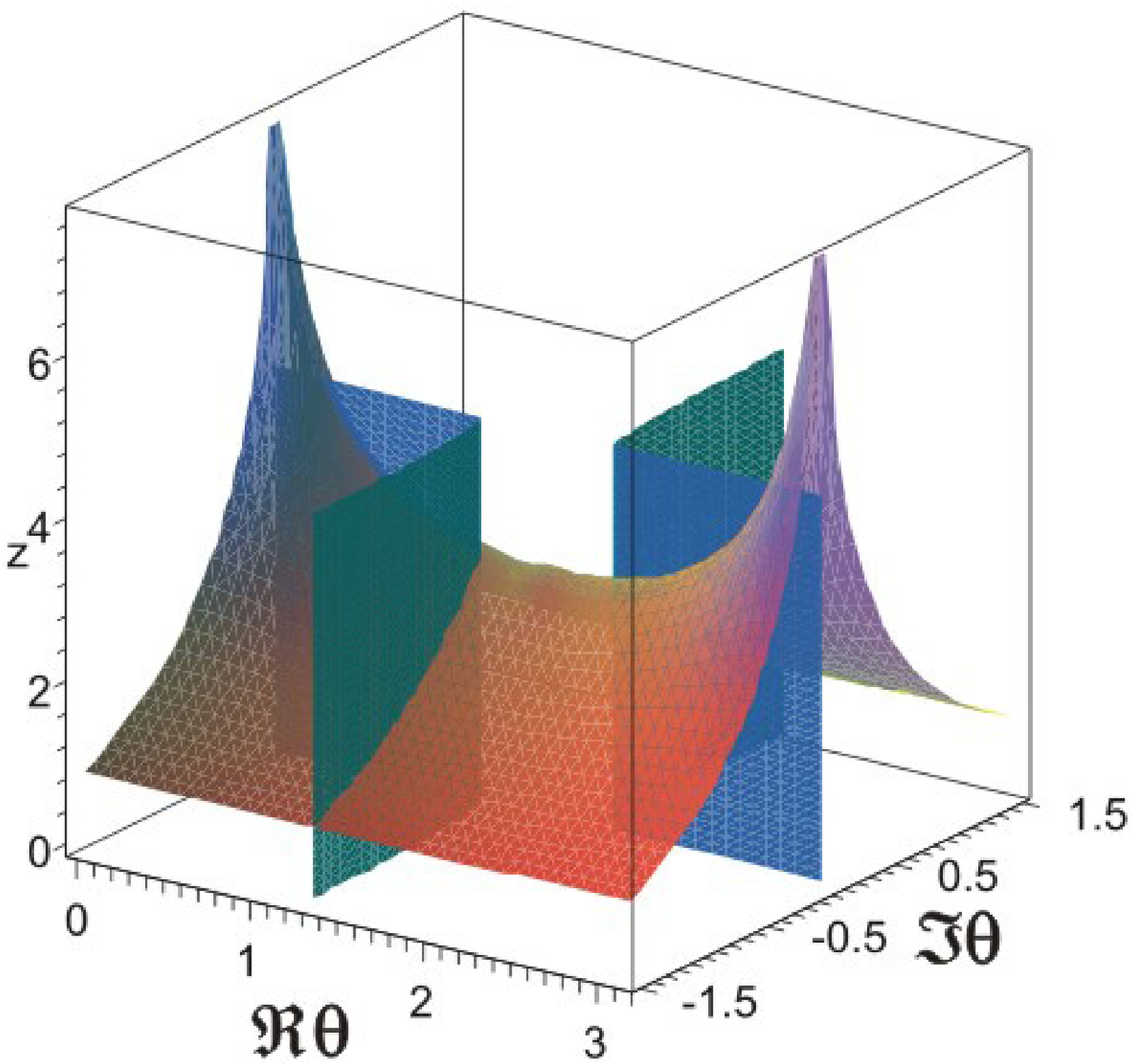}}
\caption{The evolution time $\tau$ is depicted as a line of intersection of the surfaces $z=|\Psi|$ and $\arg \Psi=0$. Left panel: $\Omega=1,\,\chi= \pi$. Right panel: $\Omega=1+0.25 i, \, \chi= \pi + 0.25 i$. The eigenvalue constraint is given by $\Omega = \rm const$. \label{ET4}}
\end{center}
%\end{minipage}
\end{figure}

\begin{figure}[tbh]
%\begin{minipage}[]{8.5cm}
\begin{center}
\scalebox{0.325}{\includegraphics{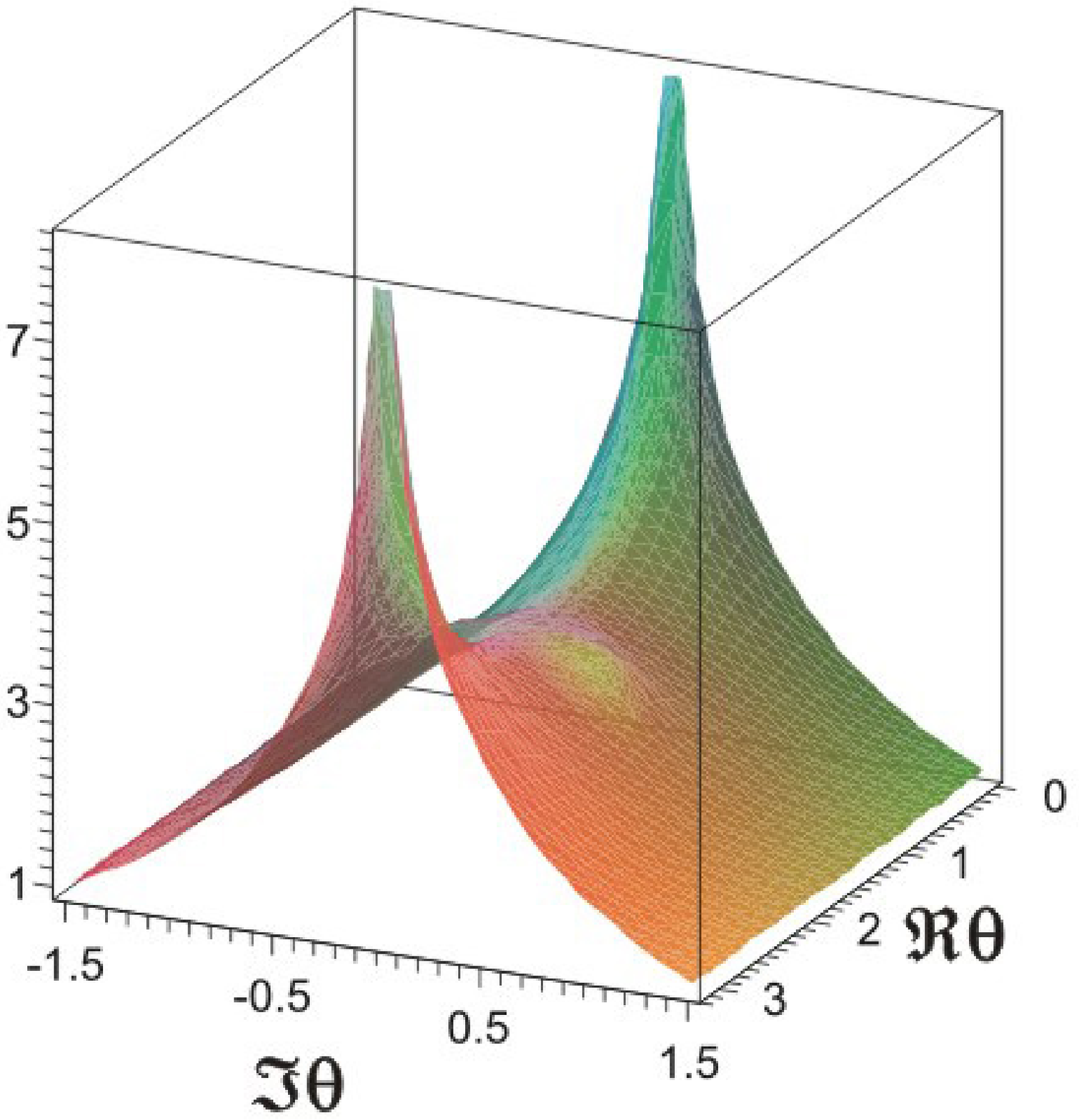}}
\scalebox{0.325}{\includegraphics{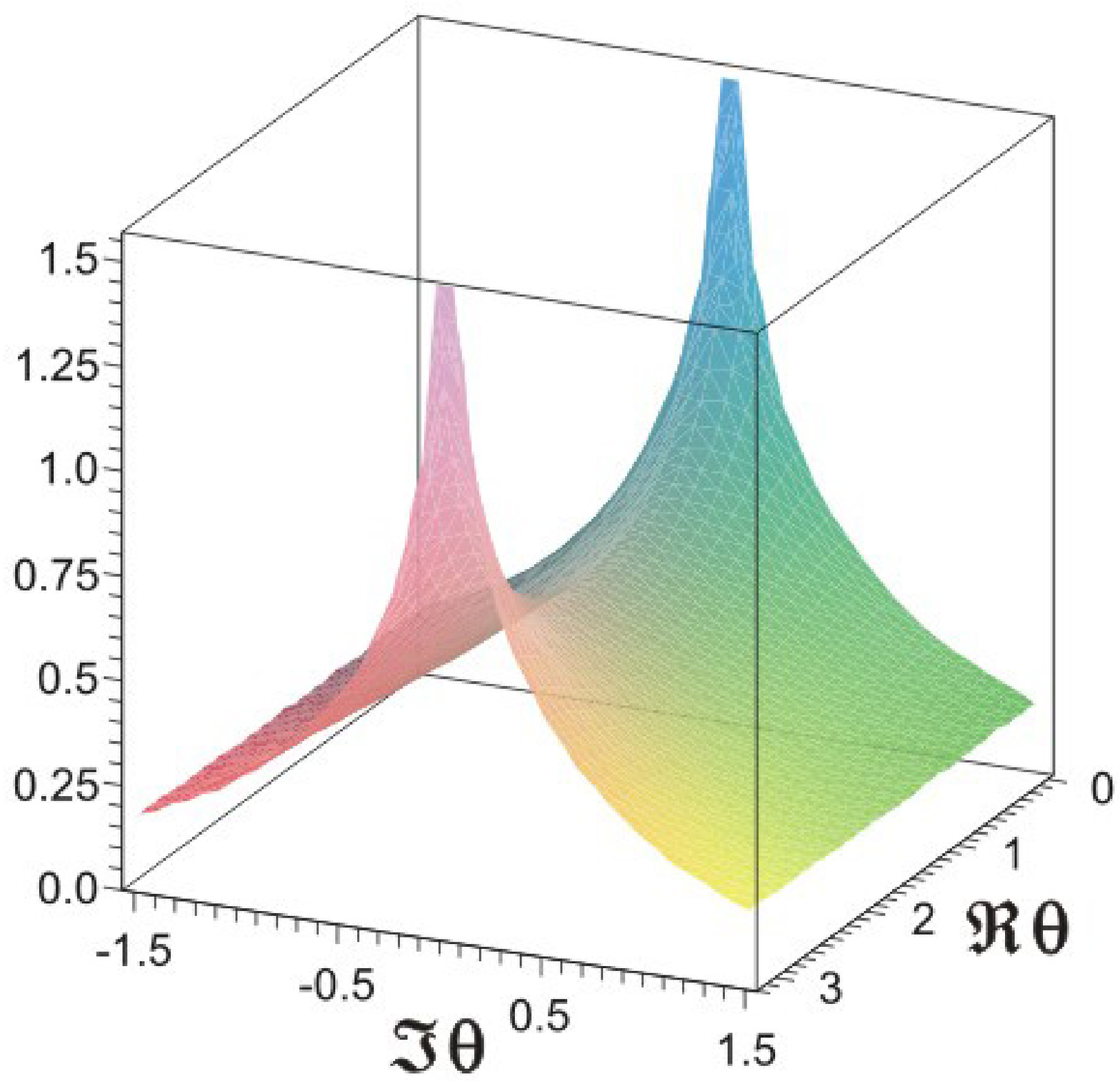}}
\caption{Plot of evolution time $\tau$ vs. $\Re \theta$ and $\Im \theta$. Left panel: $\Re\Omega=1$, $ \Im\Omega=0.1$, $\chi= \pi +i$. Right panel: $\Re\Omega=1, \Im\Omega=5, \chi= \pi$. The constraint is taken as $|\Omega| = \rm const$.\label{ET3}}
\end{center}
%\end{minipage}
\end{figure}

Further study of the critical points shows that there is no solution with a finite value of $|\theta|$ yielding the minimum of the evolution time. Indeed, when $\arg\Omega = \arg\chi$, the first solution, $\theta= \pi/2$, is related to the saddle point for the function $\tau = \tau(\theta)$, and the second one yields $\tau \rightarrow 0$ while $\Im\theta \rightarrow  \pm\infty$ (see Fig. \ref{ET4}). Thus, for a quantum mechanical system governed by a non-Hermitian Hamiltonian, the evolution time $\tau$ has a zero lower bound and may be taken to be arbitrarily small. Moreover, since, for any finite value of $|\theta|$, the minimum of $\tau$ does not exist, the non-Hermitian Hamiltonian cannot be optimized. However, for a quantum mechanical system governed by a Hermitian Hamiltonian, the latter can be optimized. Indeed, in this case, we have $\Im \theta =0$, and, hence, the saddle point becomes the point of a local minimum (Figs. \ref{ET4}, \ref{ET3} ). This agrees completely with the results obtained in \cite{CHKO,BDHD}.

Applying (\ref{PT1b}) to eliminate $\varphi$ from the effective Hamiltonian (\ref{eqH2}), we obtain $H_{ef} =H_{ef} (\Omega, \theta, \chi,\gamma)$. In the explicit form, we have
\begin{align}
\label{eqH3a}
 H_{ef}= {\lambda_0}{1\hspace{-.15cm}1} + \frac{ \Omega \cos\theta }{2}\left(
  \begin{array}{cc}
     1 & e^{-i\gamma}\big( \tan\frac{\chi}{2} - i\sqrt{\tan^2\theta -\tan^2\frac{\chi}{2} }\big) \\
  e^{i\gamma}\big( \tan\frac{\chi}{2} + i\sqrt{\tan^2\theta -\tan^2\frac{\chi}{2} } \big) &  - 1\\
  \end{array}
\right). \nonumber \\
\end{align}

To obtain the Hamiltonian in terms of initial and final states, we need to solve Eq. (\ref{B2}) and determine $|\psi_0\rangle$. After a lengthy calculation using the explicit solution of the Schr\"odinger equation (\ref{S2}), we obtain:
\begin{align}\label{eqH7}
&a =\Big (\cos \frac{\Omega\tau}{2}- i\cos\theta\sin \frac{\Omega\tau}{2}\Big )e^{-i\lambda_0\tau}, \;& b=-i e^{-i\lambda_0\tau}e^{i\varphi}\sin\theta \sin \frac{\Omega\tau}{2},\\
&\tilde a =\Big (\cos \frac{\Omega\tau}{2}+ i\cos\theta\sin \frac{\Omega\tau}{2} \Big )e^{i\lambda_0\tau}, \;& \tilde b=i e^{i\lambda_0\tau}e^{-i\varphi}\sin\theta \sin \frac{\Omega\tau}{2}. \label{eqH8}
\end{align}
Then, applying (\ref{B2a}), (\ref{PT1b}) - (\ref{PT5}) and (\ref{eqH7}), (\ref{eqH8}), we obtain
\begin{align}
\label{eqH1a}
 H=&{\lambda_0}{1\hspace{-.15cm}1}
+\frac{\Omega}{2\sin\theta\sin^2\frac{\chi}{2}}\bigg(\Big(\cos\theta\sqrt{\cos^2 \frac{\chi}{2}-\cos^2\theta } + i\sin\frac{\chi}{2}\Big)|\psi_f\rangle\langle\widetilde\psi_i| \nonumber \\
 &+\Big(\cos\theta\sqrt{\cos^2 \frac{\chi}{2}-\cos^2\theta } - i\sin\frac{\chi}{2}\Big)|\psi_i\rangle\langle\widetilde\psi_f|\bigg)
 -\frac{\Omega\cos\theta}{2\sin^2\frac{\chi}{2}}
\Big(|\psi_i\rangle\langle\widetilde\psi_i| + |\psi_f\rangle\langle\widetilde\psi_f|\Big), \nonumber \\
\end{align}
where $\cos^2 \frac{\chi}{2}= \langle\widetilde\psi_f|\psi_i\rangle \langle\widetilde\psi_i|\psi_f\rangle$.

The generic optimization problem on the complex Bloch sphere is, for a given  final vector $\mathbf n_f = (\sin\chi \cos\gamma, \sin\chi \sin\gamma, \cos\chi) $, to determine the complex angle $\theta$ for which, under a given set of eigenvalue constraints, the initial vector $\mathbf n_i =(0,0,1)$ evolves into the final vector $\mathbf n_f$ in the smallest time $\tau$. As follows from the preceding analysis, the non-Hermitian Hamiltonian can be optimized only if $\Im \theta =0$. In particular, it emerges that the Hermitian Hamiltonian is a special case of the latter when, in addition, $\Im\gamma = \Im \chi =0$.

As can be observed in Figs. \ref{ET4}, \ref{ET3}, the point $\theta = \pi/2$ is a saddle point for the evolution time $\tau=\tau(x,y)$, where $x=\Re \theta$ and $y= \Im \theta$. For $\theta =\pi/2$, the straightforward computation yields $\tau = |\chi/\Omega|$, and, in addition, we have $\arg \Omega= \arg\chi$. This relationship implies that, for a given $\Omega$, only final states $\mathbf n_f$ satisfying $\arg\chi=\arg \Omega$ can be reached from the initial state $\mathbf n_i$. Inserting $\theta = \pi/2$ into (\ref{eqH3a}), we obtain the effective Hamiltonian as
\begin{equation}\label{eqH5}
H_{ef}= {\lambda_0} {1\hspace{-.15cm}1}+ \frac{\Omega}{2}
\left(
  \begin{array}{cc}
    0& - ie^{-i\gamma}\\
   i e^{i\gamma} &  0\\
  \end{array}
\right).
\end{equation}
Finally, from Eq.(\ref{eqH1a}), the ``optimal'' Hamiltonian is found to be
\begin{align}
\label{eqH1c}
H=&{\lambda_0}{1\hspace{-.15cm}1}+ \frac{i\Omega}{2\sqrt{1- \langle\widetilde\psi_f|\psi_i\rangle \langle\widetilde\psi_i|\psi_f\rangle}}\bigg(|\psi_f\rangle\langle\widetilde\psi_i| -|\psi_i\rangle\langle\widetilde\psi_f|\Big ),
\end{align}
To compare with Hermitian quantum evolution, let us note that, for Hermitian quantum systems, all variables $(\chi, \theta, \varphi)$ and $\Omega$ are real. In that case, the point $\theta=\pi/2$ is the point of the local minimum, and the Hamiltonian (\ref{eqH1c}), indeed, becomes the optimal Hamiltonian. This is in accordance with the results obtained in the previous works of Carlini {\em et al} and Brody and Hook \cite{CHKO,BDHD}.

In what follows, we consider some illustrative examples, starting with the generalization of the results obtained by Carlini {\em et al} in their paper \cite{CHKO} on the quantum brachistochrone problem. In \cite{CHKO}, the Hamiltonian constraint has been imposed on the standard deviation $\Delta H$ of the Hamiltonian. In the standard normalized state, $|\psi\rangle$, we have $(\Delta H)^2 = \langle  \psi| H^2|\psi\rangle - (\langle  \psi| H|\psi\rangle)^2 = \rm const$.
\begin{figure}[tbh]
%\begin{minipage}[]{8.5cm}
\begin{center}
\scalebox{0.375}{\includegraphics{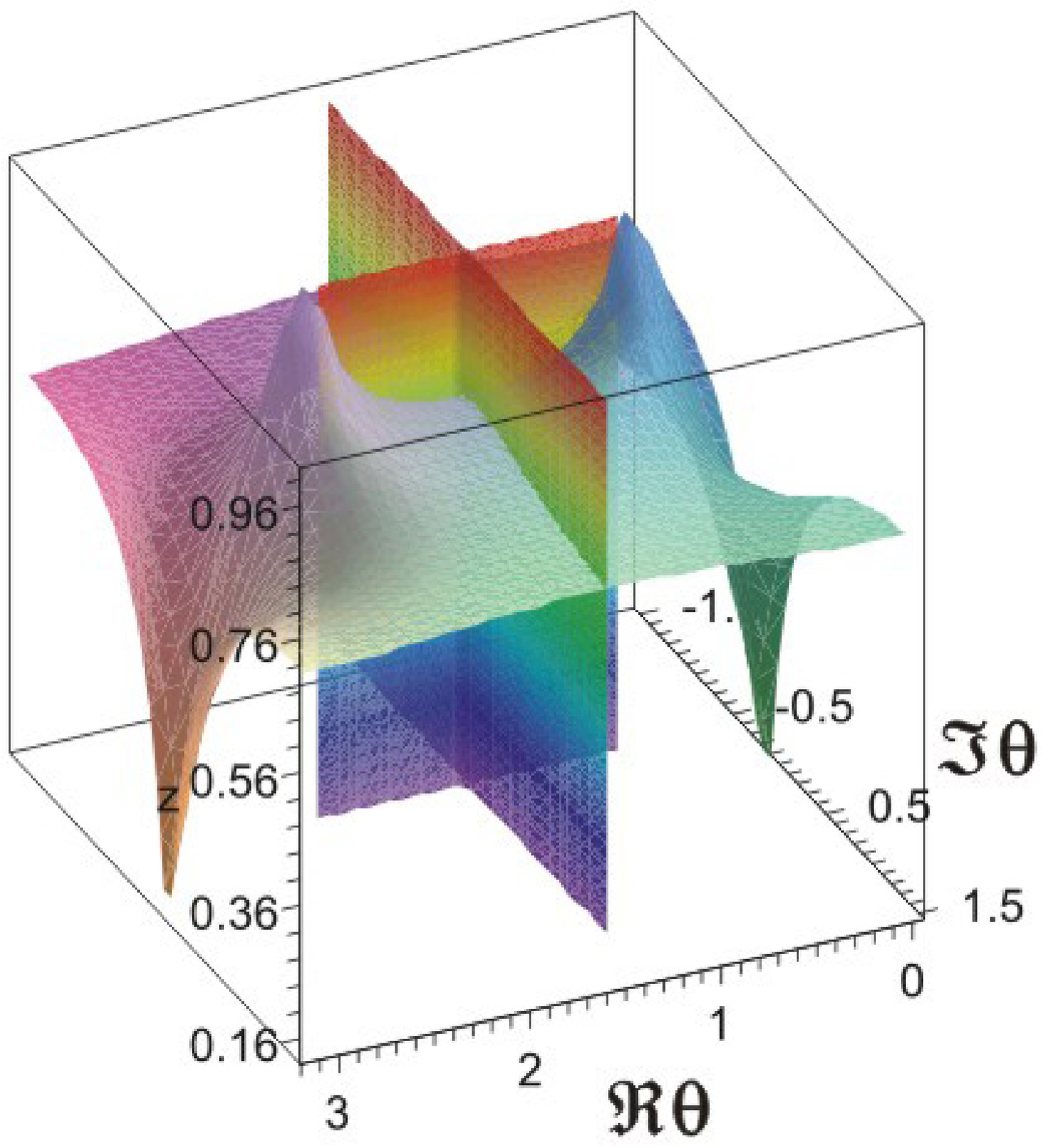}}
\hspace{-1cm}
\scalebox{0.375}{\includegraphics{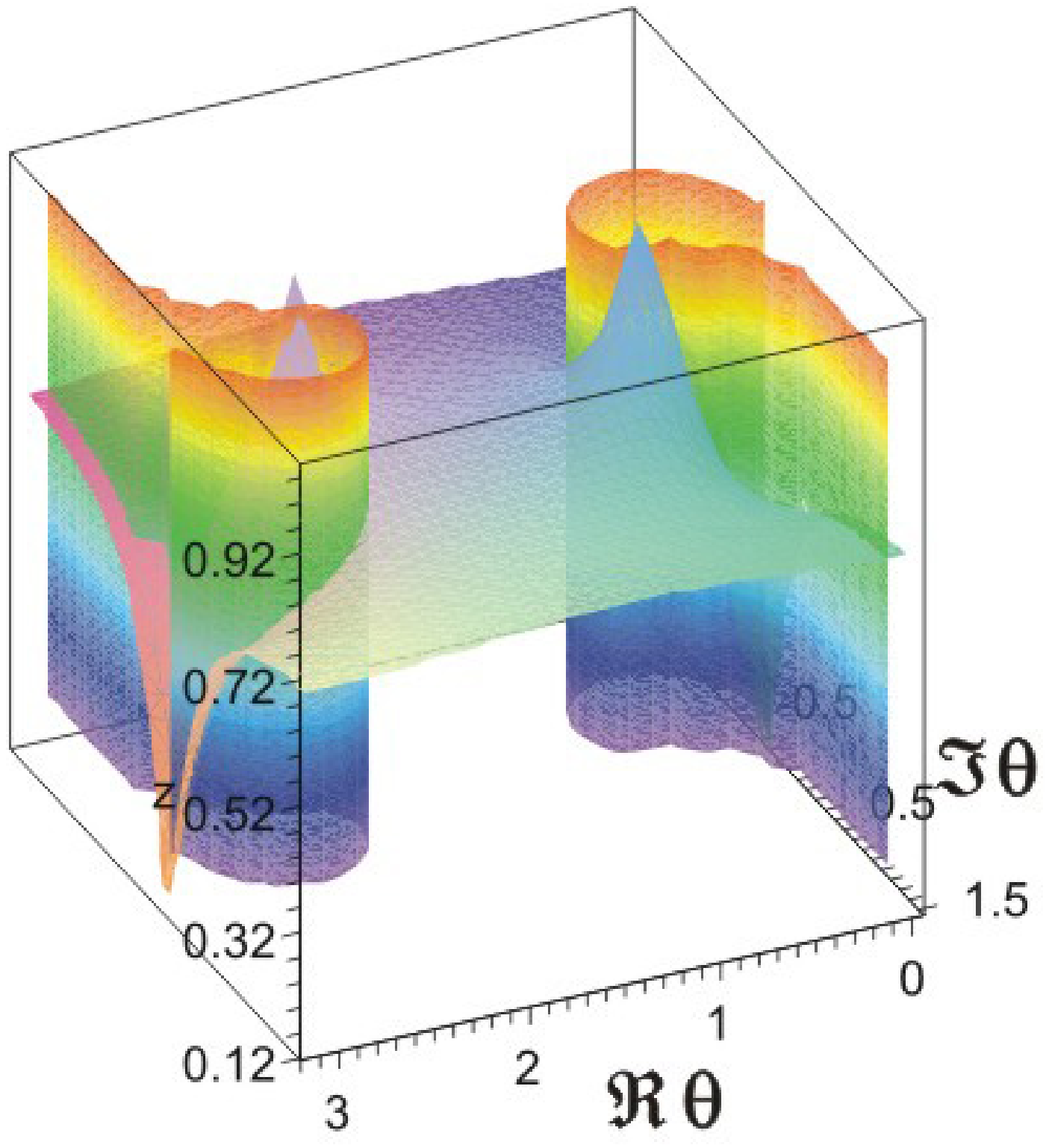}}
\caption{The evolution time $\tau$ is depicted as a line of intersection of the surfaces $z=|\tilde\Psi|$ and $\arg \tilde\Psi=0$. Left panel: $\Omega=1,\,\chi= \pi/2$. Right panel: $\Omega=1+0.25 i, \, \chi= \pi/2 + 0.25 i$. The eigenvalue constraint is given by $\Delta E = 1$. \label{ET4a}}
\end{center}
%\end{minipage}
\end{figure}
The natural generalization of the complex energy variance to the non-Hermitian Hamiltonian is given by $(\Delta E)^2 = \langle \tilde \psi| H^2|\psi\rangle - (\langle \tilde \psi| H|\psi\rangle)^2$, where $\langle \tilde \psi|\psi\rangle =1$. Then, applying this to the Hamiltonian (\ref{eqH3}) and with the help of Eq. (\ref{eq1}), we obtain $\Delta E = (\Omega \sin\theta)/2$.

Imposing the Hamiltonian constraint as $\Delta E = \rm const$, from Eq. ({\ref{PT5}}) we obtain
\begin{align}
\tan \frac{\tau\Delta E}{\sin\theta}=
\frac{\sin\frac{\chi}{2}}{\sqrt{\cos^2\frac{\chi}{2} - \cos^2\theta}}
\label{PT5b}\
\end{align}
This yields $\tau = |\tilde \Psi|$, where
\begin{align}
\label{PT6a}
\tilde\Psi = \frac{\sin\theta}{\Delta E} \arctan\Bigg(\frac{\sin\frac{\chi}{2}}{\sqrt{\cos^2\frac{\chi}{2} - \cos^2\theta}} \Bigg)
\end{align}
In addition, since $\tau$ is a real function, one should impose the following constraint $\arg\tilde \Psi =0$. In Fig. \ref{ET4a}, the evolution time $\tau$ is depicted as the line of intersection of the surfaces $z= |\tilde\Psi|$ and $\arg\tilde \Psi =0$.

The following example is related to the recent controversy on the possibility of achieving faster evolution in a quantum mechanical system governed by a non-Hermitian $\cal PT$-symmetric Hamiltonian as compared to the equivalent Hermitian system \cite{BBJ,BBJM,AF,MA}. Note that the critique of the results obtained in Ref. \cite{BBJ} is essentially based on the following theorem \cite{MA}: {\em The lower bound on the travel time (upper bound on the speed) of unitary evolutions is a universal quantity, independent of whether the evolution is generated by a Hermitian or non-Hermitian Hamiltonian}. Analyzing the proof of this theorem, one can see that it is based on following assumption: the minimal travel time is realized by quantum evolution along the geodesic path in the Hilbert space joining initial and final states. This is true in the case of the Hermitian Hamiltonian; however, as we further show, for a non-Hermitian Hamiltonian, the situation is quite different.

Without loss of generality, we can further confine our attention to the case of $\Im\chi= 0$. Taking into consideration that eigenvalues of $\cal PT$-symmetric Hamiltonian are real, and employing Eq. (\ref{PT1a}), we find $\Re\theta = \pi/2$ and $\Re(\varphi - \gamma) = \pm \pi/2$. In what follows, to be definite, we will choose the upper sign. Then, denoting the imaginary part of $\theta$ as $\eta$ and substituting  $\theta= \pi/2 + i \eta$ into Eq. (\ref{eqH3}), we obtain
\begin{equation}\label{eqH3b}
H_{ef}= {\lambda_0} {1\hspace{-.15cm}1}+ \frac{1}{2}\,{\mathbf  \Omega}\cdot \boldsymbol \sigma
\end{equation}
where $\mathbf \Omega=\Omega(\cosh\eta \cos\varphi, \cosh\eta \sin\varphi, -i\sinh\eta)$. Further, we assume $\Omega \geq 0$.

Using a convenient parametrization,
\begin{equation}
   x=\Omega\cosh\eta \cos\varphi,\;y=\Omega \cosh\eta \sin\varphi, \; z=\Omega\sinh\eta
\end{equation}
we have ${x^2+y^2-z^2}= \Omega^2$.  Thus, the eigenvalue constraint $\Omega = \rm const$ defines a one-sheeted hyperboloid embedded in a flat three-dimensional space $\mathbb R^3$. Finally, inserting $\theta= \pi/2 + i \eta$ into Eq. (\ref{eq1}), we obtain
\begin{align}
&\mathbf n(t) = \left(\begin{array}{l}
-i\sinh\eta \cosh\eta(1-\cos\Omega t)\cos\varphi + \cosh\eta \sin\Omega t \sin\varphi\\
-i\sinh\eta \cosh\eta(1-\cos\Omega t)\sin\varphi-\cosh\eta \sin\Omega t\cos\varphi \\
\cosh^2 \eta \cos\Omega t - \sinh^2 \eta
\end{array}\right)
 \label{B3}
\end{align}
and, evidently, $\mathbf n(0)= (0,0,1)$.

We show further that, for a given $\Omega$, the evolution of a quantum mechanical system governed by the non-Hermitian Hamiltonian with real spectra can be considered as motion over the surface of the unit one-sheeted hyperboloid $\mathbb H^2$. Let $\mathbf n(t)$ be the Bloch vector given by (\ref{B3}).
Then, we define the map $\mathbf n(t) \rightarrow \mathbf m(t)$ as
\begin{align}
\label{B5a}
& m^1 = n^3 = \cosh^2 \eta \cos\Omega t - \sinh^2 \eta \\
& m^2 = n^1 \sin\varphi - n^2\cos\varphi = \cosh\eta \sin\Omega t \\
& m^3 = i(n^1 \cos\varphi + n^2\sin\varphi)= \sinh\eta \cosh\eta(1-\cos\Omega t)
 \label{B5}
\end{align}
This yields $g_{ij}\,m^i m^j =1$, where $g_{ij}= \rm diag(1,1,-1)$ is the indefinite metric, and, hence, Eqs. (\ref{B5a}) - (\ref{B5}) define the map $S^2_c \rightarrow {\mathbb H}^2$. Thus, the vector $\mathbf m(t)$ traces out a trajectory on the unit one-sheeted hyperboloid $\mathbb H^2$, while the quantum mechanical system evolves in the Hilbert space. From (\ref{B5a}) -- (\ref{B5}), it then follows that the initial Bloch vector $\mathbf n_i = (0,0,1)$  maps to $\mathbf m_i = (1,0,0)$, and, for the final state defined by $\mathbf n_f=(\sin\chi \cos\gamma, \sin\chi \sin\gamma, \cos\chi)$, we have
\begin{align}
\mathbf n_f \rightarrow \mathbf m_f = \big(\cos\chi,\sqrt{\sin^2\chi + \tanh^2\eta(1- \cos\chi)^2 },\tanh\eta(1- \cos\chi) \big)
\label{PT10}
\end{align}

\begin{figure}[tbh]
%\begin{minipage}[]{8.5cm}
\begin{center}
\scalebox{0.375}{\includegraphics{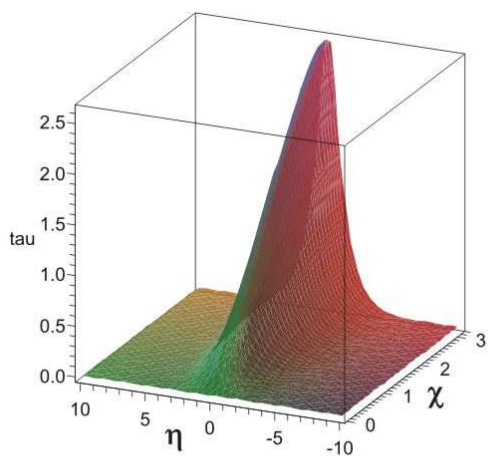}}
%\scalebox{0.275}{\includegraphics{tau_1_0_Pi_0b1}}
\caption{Plot of evolution time $\tau$ vs. $\chi$ and $\eta= \Im \theta$ for a non-Hermitian Hamiltonian with real spectra ($\Omega=1$). As can be observed, for a given $\chi$, the evolution time has a maximum at the point $\eta=0$. \label{ET3a}}
\end{center}
%\end{minipage}
\end{figure}

The amount of time $\tau$ required to evolve the initial state $\mathbf m_i$ into the final state $\mathbf m_f$ can be found from Eq. (\ref{PT5}) by substituting $\theta= \pi/2 + i \eta$. The computation yields
\begin{align}\label{PT3}
 \tau=\frac{2}{\Omega} \arctan\frac{\sin\frac{\chi}{2}}{\sqrt{\cos^2\frac{\chi}{2} + \sinh^2\eta}}
\end{align}
Therefore, it follows that a non-Hermitian Hamiltonian with real spectra $\tau$ has the upper bound $\tau \leq \chi/\Omega$, where $\chi/\Omega$ is the evolution time for the optimal Hermitian Hamiltonian (\ref{eqH1c}) and can be obtained from (\ref{PT3}) taking $\eta =0$ (see Fig. \ref{ET3a}).
Our results are in agreement with those obtained previously by Bender {\em et al} in \cite{BBJ}, and they contradict the conclusions on the impossibility of achieving a faster evolution for non-Hermitian quantum evolution than for Hermitian evolution \cite{MA}.

To compare $\tau$ with the geodesic evolution time $\tau_g$, we must consider the induced intrinsic metric $ds^2 = g_{ij}\,dm^i d m^j$ on the unit one-sheeted hyperboloid ${\mathbb H}^2$, where $g_{ij}= \rm diag(1,1,-1)$. Let $\mathbf m $ be a unit vector with respect to the indefinite metric on $g_{ij}$. Using a convenient parametrization,
\begin{align}\label{PR}
m^1 = \cosh \rho \cos \nu, \, m^2= \cosh \rho \sin \nu, \, m^3= \sinh \rho
\end{align}
where $-\infty <\rho <\infty $ and $0 \leq\nu < 2\pi$, we find
\begin{eqnarray}\label{M1}
ds^2  =\cosh^2 \rho \,d\nu^2 -d\rho^2
\end{eqnarray}
Any geodesic on $\mathbb H^2$ can be obtained as a solution of the geodesic equation
\begin{eqnarray}\label{G2}
\frac{d^2 x^i }{ds^2} + \Gamma^i_{jk}\frac{d x^j}{ds}\frac{d x^k}{ds} = 0
\end{eqnarray}
For the metric (\ref{M1}), it takes the form
\begin{eqnarray}\label{G1}
&&\frac{d^2 \rho }{ds^2} + \sinh\rho  \cosh\rho \, \bigg(\frac{d\nu}{ds}\bigg)^2 = 0 \\
&&\frac{d^2 \nu }{ds^2} + 2\tanh\rho\, \frac{d\rho}{ds} \frac{d\nu}{ds} = 0
\label{G1a}
\end{eqnarray}
We note that, setting $\mathbf m =(\cosh \rho \cos \nu,\cosh \rho \sin \nu,\sinh \rho)$, one can rewrite the geodesic equation as
\begin{eqnarray}\label{G4}
\frac{d^2 \mathbf m}{ds^2} + \mathbf m = 0
\end{eqnarray}
Its solution is given by
\begin{eqnarray}\label{G2b}
 \mathbf m (s) = {\cos s} \;\mathbf m_i  + {\sin s} \;\frac{d \mathbf m_i}{ds}
\end{eqnarray}
Taking the initial state as $ \mathbf m_i = (1,0,0)$ and the final state as $ \mathbf m_f$ from Eq. (\ref{PT10}), we have
\begin{align}\label{PT9}
\mathbf m= \Big(\cos s , \sqrt{1 + \tanh^2\eta \tanh^2\frac{\chi}{2}}\,\sin s,
\tanh\eta \tanh\frac{\chi}{2}\, \sin s\Big)
\end{align}
Now, comparing (\ref{PT9}) with Eqs.(\ref{B5a}) -- (\ref{B5}), we conclude that $\mathbf m(t)$ determined by the Schr\"odinger equation yields the geodesic evolution only if $\eta =0$. In addition, for the geodesic motion, we have $s = \Omega t$. Using these results, we can rewrite (\ref{PT3}) as
\begin{align}\label{PT8}
 \tau= \tau_g \frac{2}{\chi} \arctan\frac{\sin\frac{\chi}{2}}{\sqrt{\cos^2\frac{\chi}{2} + \sinh^2\eta}}
\end{align}
where $\tau_g =\chi/\Omega$ is the geodesic evolution time. This implies that $\tau < \tau_g$ if $\eta \neq 0$, and $\tau = \tau_g$ if $\eta =0$. Furthermore, whereas, for $\theta = \pi/2$ ($\Im \theta=\eta =0$), the optimal Hermitian Hamiltonian provides the shortest travel time, for a non-Hermitian Hamiltonian, the geodesic motion over the one-sheeted hyperboloid yields only the upper bound on the evolution time.

\begin{figure}[tbh]
%\begin{minipage}[]{8.5cm}
\begin{center}
\scalebox{0.375}{\includegraphics{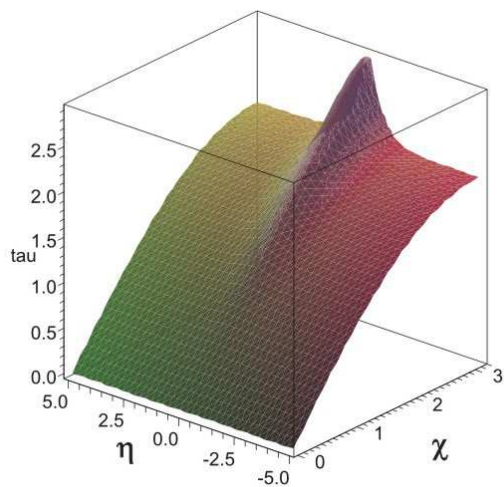}}
\caption{Plot of $L$ vs. $\chi$ and $\eta$ for the quantum evolution generated by the non-Hermitian Hamiltonian (\ref{eqH3b}). \label{LS1}}
\end{center}
%\end{minipage}
\end{figure}

Computing the distance on $\mathbb H^2$ between the initial $\mathbf m_i$ and final $\mathbf m_f $ states, calculated along the curve $\mathcal C$  defined by Eqs. (\ref{B5a}) -- (\ref{B5}), we obtain
\begin{align}\label{L3}
L = \int_{\mathcal C}\sqrt{g_{ij}\dot m^i \dot m^j} \,dt = \tau \Omega \cosh\eta = 2\cosh\eta \arctan\Bigg(\frac{\sin\frac{\chi}{2}}{\sqrt{\cos^2\frac{\chi}{2} + \sinh^2\eta}} \Bigg)
\end{align}
As can be easily shown, $L$ is bounded as $2 \sin(\chi/2) < L \leq \chi$, where the upper bound is determined by the geodesic distance $L_g=\Omega \tau_g = \chi$ between the initial and final states on $\mathbb H^2$ (Fig. \ref{LS1}). The lower bound, $L \rightarrow 2\sin(\chi/2)$, is reached while $\eta \rightarrow \infty$. Note that the geodesic distance between $\mathbf m_i$ and $\mathbf m_f$ calculated on on $\mathbb H^2$ with the help of the indefinite metric (\ref{M1}) coincides with the geodesic distance between the related initial  $\mathbf n_i$ and final  $\mathbf n_f$ states on the conventional Bloch sphere $S^2$. However, in the latter case, the system is governed by the optimal Hermitian Hamiltonian, and $L_g=\chi$ is the shortest distance on $S^2$ between $\mathbf n_i$ and $\mathbf n_f$.

To compare the evolution time for the non-Hermitian and Hermitian Hamiltonians and conclude which one is faster, one should not only fix the geodesic distance between initial and final states, as was pointed out in \cite{MA}, but also impose the same set of constraints for both cases. Here, following \cite{CHKO}, we assume that the Hamiltonian constraint $2\Delta E = \Omega \cosh \eta = \rm const$ is held. Then, denoting $\Omega\cosh\eta$ as $\omega$, we obtain from Eq. (\ref{PT8}) the following result:
\begin{align}\label{PT8a}
 \tau= \frac{2}{\Omega} \arctan\frac{\Omega\sin\frac{\chi}{2}}{\sqrt{\omega^2 -\Omega ^2\sin^2\frac{\chi}{2} }}
\end{align}
This yields
\begin{align}\label{PT8b}
\frac{2}{\omega}\sin\frac{\chi}{2}\leq \tau \leq \frac{\chi}{\omega}
\end{align}
As can be seen, the evolution time $\tau$ reaches it maximum $\tau_{max} = \chi/\omega$ when $\Omega = \omega$. This implies $\eta =0$, and we have the quantum mechanical evolution governed by the optimal Hermitian Hamiltonian for which $\theta = \pi/2$. The lower bound, $\tau_{min}= ({2}/{\omega})\sin({\chi}/{2})$, is obtained at the exceptional point defined by the condition $\Omega =0$. Similar consideration of the distance (\ref{L3}) between $\mathbf m_i$ and $\mathbf m_f$ yields
\begin{align}\label{L3a}
L = \omega \tau   =\frac{2\omega}{\Omega} \arctan\frac{\Omega\sin\frac{\chi}{2}}{\sqrt{\omega^2 -\Omega ^2\sin^2\frac{\chi}{2} }}
\end{align}
and
\begin{align}\label{PT8c}
{2}\sin\frac{\chi}{2}\leq L \leq {\chi}
\end{align}
The upper bound $L_{max} = \chi$, being identical to the geodesic distance between the initial and final states defined either on the Bloch sphere $S^2$ or one-sheeted hyperboloid $\mathbb H^2$, is achieved for the Hermitian Hamiltonian. The lower bound, $L_{min}= 2\sin({\chi}/{2})$, is obtained at the exceptional point.

Thus, as has been shown above, the geodesic distance $L_g$ between $\mathbf n_i$ and $\mathbf n_f$ calculated over the Bloch sphere $S^2$ is identical to the geodesic distance between  $\mathbf m_i$ and  $\mathbf m_f$ computed over the one-sheeted hyperboloid $\mathbb H^2$. Moreover, the amount of time $\tau_g$ required to evolve $\mathbf n_i$ into $\mathbf n_f$ on the Bloch sphere and that required to evolve $\mathbf m_i$ into  $\mathbf m_f$ on $\mathbb H^2$ by the geodesic evolution, is the same. This is in accordance with the conclusions made in \cite{MA}. However, in the case of the Hermitian Hamiltonian, $\tau_g$ is the lower bound on the evolution time, and, for the non-Hermitian Hamiltonian, it yields only the upper bound on the evolution time. Furthermore, as follows from Eq. (\ref{L3}) for the non-Hermitian Hamiltonian, the evolution speed $v= ds/dt$ is given by $v= \omega =\Omega\cosh\eta$ (we recall that $\eta =\Im \theta$). Hence, $v \geq v_g$, where $v_g= \Omega$ is the speed of the geodesic evolution. Similar consideration of the quantum mechanical system governed by the Hermitian Hamiltonian yields $v= \Omega\sin\theta$, and, obviously, $v \leq v_g$. This proves that, indeed, non-Hermitian quantum mechanics can be faster than Hermitian evolution \cite{BBJ}.

In summary, we have formulated a geometric problem on the complex Bloch sphere to find the optimal non-Hermitian Hamiltonian and the optimal time evolution for a given pair of initial and final states and eigenvalue constraints. In contrast to the Hermitian quantum system, generic non-Hermitian Hamiltonians generate non-unitary transformations $|\psi_i\rangle \rightarrow |\psi_f\rangle$ such that the evolution may be realized in an arbitrarily short time. This is in agreement with the previous results obtained for $\cal PT$-symmetric quantum systems and some specific non-Hermitian Hamiltonians \cite{BBJ,AF}.

\section*{Acknowledgements}
 This work has been supported by research grants
SEP-PROMEP 103.5/04/1911 and CONACyT U45704-F.

%\bibliography{EP}
\bibliography{bs_arxiv}

\begin{thebibliography}{10}
\expandafter\ifx\csname url\endcsname\relax
  \def\url#1{\texttt{#1}}\fi
\expandafter\ifx\csname urlprefix\endcsname\relax\def\urlprefix{URL }\fi

\bibitem{CHKO}
A.~Carlini, A.~Hosoya, T.~Koike, Y.~Okudaira, Time-optimal quantum evolution,
  Phys. Rev. Lett. 96~(6) (2006) 060503.

\bibitem{BDHD}
D.~C. Brody, D.~W. Hook, {On optimum Hamiltonians for state transformations},
  J. Phys. A 39~(11) (2006) L167--L170.

\bibitem{BDC}
D.~C. Brody, Elementary derivation for passage times, J. Phys. A 36~(20) (2003)
  5587--5593.

\bibitem{BBJ}
C.~M. Bender, D.~C. Brody, H.~F. Jones, B.~K. Meister, {Faster than Hermitian
  Quantum Mechanics}, Phys. Rev. Lett. 98~(4) (2007) 040403.

\bibitem{WJ}
J.~Wong, {Results on Certain Non-Hermitian Hamiltonians}, J. Math. Phys. 8~(10)
  (1967) 2039--2042.

\bibitem{BH}
H.~C. Baker, {Non-Hermitian Dynamics of Multiphoton Ionization}, Phys. Rev.
  Lett. 50~(20) (1983) 1579--1582.

\bibitem{BHC}
H.~C. Baker, {Non-Hermitian quantum theory of multiphoton ionization}, Phys.
  Rev. A 30~(2) (1984) 773--793.

\bibitem{BS}
H.~C. Baker, R.~L. Singleton, {Non-Hermitian quantum dynamics}, Phys. Rev. A
  42~(1) (1990) 10--17.

\bibitem{TT}
T.~Tanaka, {On existence of a biorthonormal basis composed of eigenvectors of
  non-Hermitian operators}, J. Phys. A: Math. and Gen. 39~(24) (2006)
  7757--7761.

\bibitem{B}
M.~V. Berry, {Physics of nonhermitian degeneracies}, Czech. J. Phys. 54 (2004)
  1039.

\bibitem{CWZL}
S.-I. Chu, Z.-C. Wu, E.~Layton, {Density matrix formulation of complex
  geometric quantum phases in dissipative systems}, Chem. Phys. Lett. 157~(1,2)
  (1989) 151--158.

\bibitem{CT}
S.-I. Chu, D.~A. Telnov, {Beyond the Floquet theorem: generalized Floquet
  formalisms and quasienergy methods for atomic and molecular multiphoton
  processes in intense laser fields}, Phys. Rep. 390 (2004) 1--131.

\bibitem{BBJM}
C.~M. Bender, D.~C. Brody, H.~F. Jones, B.~K. Meister, {Comment on the Quantum
  Brachistochrone Problem}, arXiv:0804.3487 [quant-ph].

\bibitem{AF}
P.~E.~G. {Assis}, A.~{Fring}, {The quantum brachistochrone problem for
  non-Hermitian Hamiltonians}, {quant-ph/0703254}.

\bibitem{MA}
A.~Mostafazadeh, {Quantum Brachistochrone Problem and the Geometry of the State
  Space in Pseudo-Hermitian Quantum Mechanics}, Phys. Rev. Lett. 99~(13) (2007)
  130502.

\end{thebibliography}

\end{document}